\documentclass[doublecol]{epl2} 
\usepackage{amsmath,amssymb,latexsym,MnSymbol}
\usepackage{hyperref}
\usepackage{graphicx}
\usepackage{multirow,booktabs}
\usepackage{color}

\begin{document}
\title{Spatial correlations, clustering and percolation-like transitions in homicide crimes}

\author{L. G. A. Alves\inst{1}\!\!\!\thanks{E-mail: \email{lgaalves@dfi.uem.br}} \and E. K. Lenzi\inst{1,3} \and R. S. Mendes\inst{1} \and H. V. Ribeiro\inst{1,2}\!\!\!\thanks{E-mail: \email{hvr@dfi.uem.br}}}
\shortauthor{L. G. A. Alves \etal}

\institute{                    
\inst{1} Departamento de F\'isica, Universidade Estadual de Maring\'a, Maring\'a, PR 87020-900, Brazil\\
\inst{2} Departamento de F\'isica, Universidade Tecnol\'ogica Federal do Paran\'a, Apucarana, PR 86812-460, Brazil\\
\inst{3} Departamento de F\'isica, Universidade Estadual de Ponta Grossa, Ponta Grossa, PR 84030-900, Brazil\\
}


\abstract{
The spatial dynamics of criminal activities has been recently studied through statistical physics methods; however, models and results have been  focused on local scales (city level) and much less is known about these patterns at larger scales such as at a country level. Here we report on a characterization of the spatial dynamics of the homicide crimes along the Brazilian territory using data from all cities ($\sim$5000) in a period of more than thirty years. Our results show that the spatial correlation function in the per capita homicides decays exponentially with the distance between cities and that the characteristic correlation length displays an acute increasing trend in the latest years. We also investigate the formation of spatial clusters of cities via a percolation-like analysis, where clustering of cities and a phase transition-like behavior describing the size of the largest cluster as a function of a homicide threshold are observed. This transition-like behavior presents evolutive features characterized by an increasing in the homicide threshold (where the transitions occur) and by a decreasing in the transition magnitudes (length of the jumps in the cluster size). We believe that our work sheds new lights on the spatial patterns of criminal activities at large scales, which may contribute for better political decisions and resources allocation as well as opens new possibilities for modeling criminal activities by setting up fundamental empirical patterns at large scales.
}

\maketitle
\section{Introduction}
The study of the social phenomena is now ubiquitous in the physicist's research agenda. In particular, methods based on statistical physics have been shown very powerful when applied to social systems~\cite{Castellano,Galam, Vespignani,Conte}. An important example is related to the empirical investigations and modeling of criminal activities~\cite{Alves1,Alves2,Alves3,Bettencourt1,Bettencourt2,Bettencourt3,Gomez-Lievano,Alves4}, which have been recently reviewed by  D'Orsogna and Perc~\cite{Perc}. On a local scale (city level), a well-known pattern is related to the so-called ``broken windows theory''. Proposed by Wilson and Kelling~\cite{Wilson} in 1982 and widely accepted among criminologist, this theory can be summarized by the idea that degraded urban environments foster criminal activities in their neighborhoods. Indeed, empirical findings suggest that criminals tend to return to previously-visited locations~\cite{Short} and the violation of social norms/rules causes the spread of disorder~\cite{Keizer}. These behaviors have been modeled by reaction--diffusion equations~\cite{Short2,Short3,Rodriguez} and self-exciting point processes~\cite{Mohler,Mohler2,Lewis}. Remarkably, statistical methods based on such models have been actually employed for identifying possible crime areas in several security administration departments~\cite{Harries,Perc,Keizer}.

On larger scales, such as at a country level, much less is known about the spatial dynamics of criminal activities. Questions on whether criminal activities in a given city have influence in its neighboring cities or if criminal activities also form spatial clusters of cities have not been addressed yet. In this context, Brazil is (unfortunately) an ideal place for addressing such questions: it is one of the most violent countries in the world and it is also a country with continental dimensions. Here we investigate the spatial correlations and the clustering patterns of homicides in all Brazilian cities; we further quantify how these two aspects of the spatial dynamics of homicides have evolved along the period of 1980 to 2011. 

Our findings show that the spatial correlation function in the per capita homicides decays exponentially with the distance between cities; we further observe that the characteristic correlation length displays an acute increasing trend in the latest years ($\approx14$~km by year). These behaviors are not directly related with the spatial correlations in the population size, which present power-law decays with a quite stable exponent. We also investigate the formation of spatial clusters of cities as a function of a lower bound for the per capita homicides. The results reveal the emergence of spatial clusters and that the size of the largest cluster depends on the employed threshold in a non-trivial way. Similarly to what happens in phase transitions, the size of the largest cluster abruptly changes around a specific value of the homicide threshold. This transition-like behavior has also evolved: the homicide threshold where the transitions occur has increased, while the magnitude of the transitions (length of the jumps in the cluster size) has decreased over the years. Thus, our work supports the idea that the number of homicides of a city affects the value of this urban metric in nearby cities (tens of kilometers away) and that this correlated behavior has intensified over the last few years. In the following, we present our dataset, the analysis on the spatial correlations, the clustering analysis and, finally, some concluding remarks.

\section{Data presentation}
\begin{figure}[!ht]
\begin{center}
\includegraphics[scale=0.40]{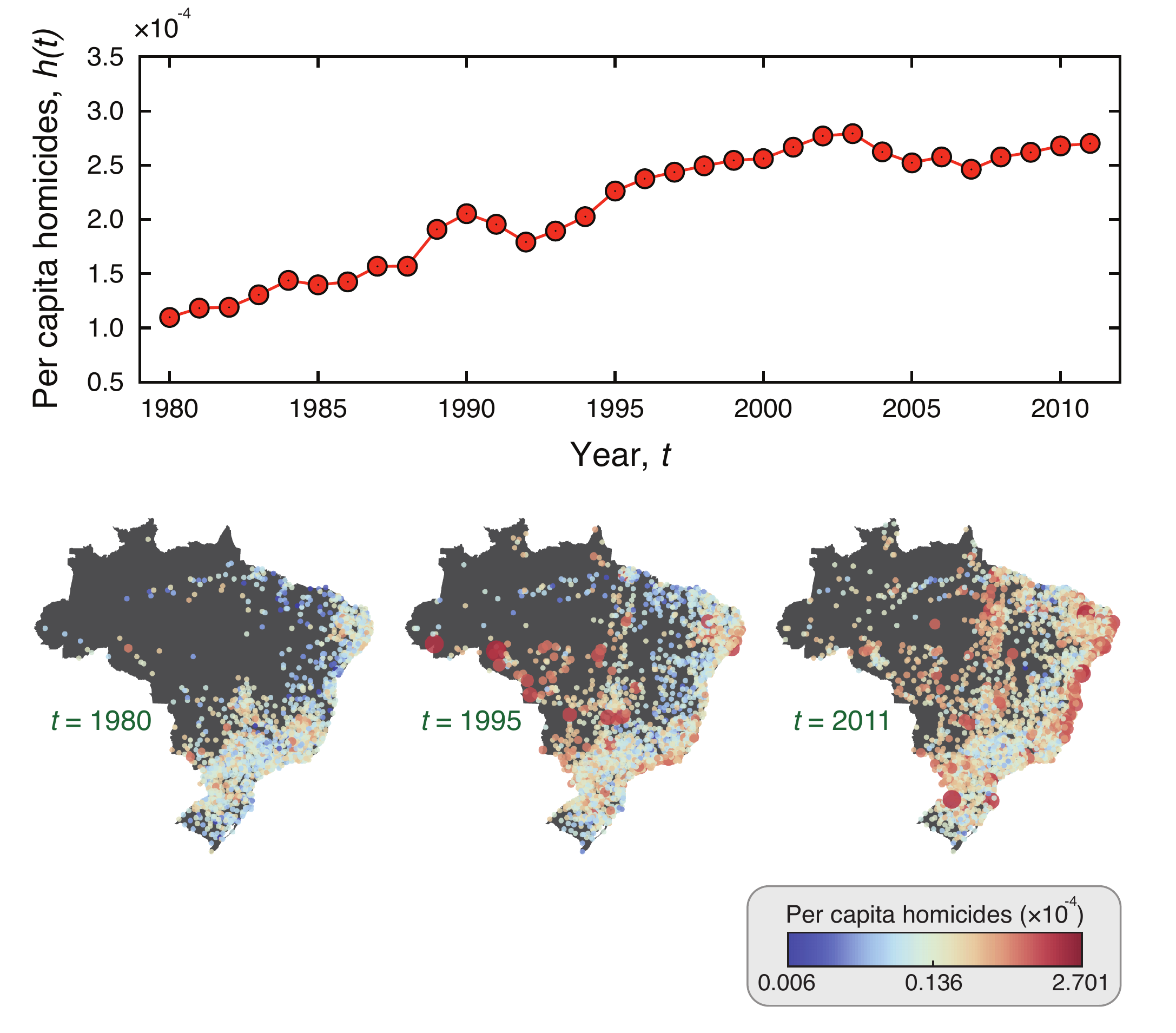}
\end{center}
\caption{Time evolution of the per capita homicide in Brazil and its spatial distribution among the Brazilian cities. The plot shows the evolution of the overall per capita homicide $h(t)$ in Brazil during the period of 1980 to 2011 and the three maps illustrate the spatial distribution of $h_i(t)$ over the Brazilian territory in the years 1980, 1995 and 2011. For the maps, each point represents a city and the color code stands for the value of $h_i(t)$, the size of point is also proportional to $h_i(t)$.}
\label{fig:1}
\end{figure}

We have accessed data of all Brazilian cities in the period of 1980 to 2011 (31 years) made freely available by the Department of Informatics of the Brazilian Public Health System~\cite{datasus}. For each city $i$ in given year $t$, we have the number of homicides $H_i(t)$, the population size $N_i(t)$ and the geographic location of the city (latitude and longitude). Figure~\ref{fig:1} illustrates our dataset by showing the evolution of the per capita homicide in Brazil $h(t)=[\sum H_i(t)]/[\sum N_i(t)]$ and the spatial distribution of the per capita homicide for each city, $h_i(t)=H_i(t)/N_i(t)$, over the Brazilian territory. We note that the per capita homicide in Brazil has more than doubled during the period of 31 years covered by our data. The maps of this figure also provide clues of spatial correlations and clustering in the spatial distribution of $h_i(t)$; specifically,  we observe the emergence of areas (group of cities) with large and small values of $h_i(t)$.

\section{Results}
\begin{figure*}[!ht]
\begin{center}
\includegraphics[scale=0.4]{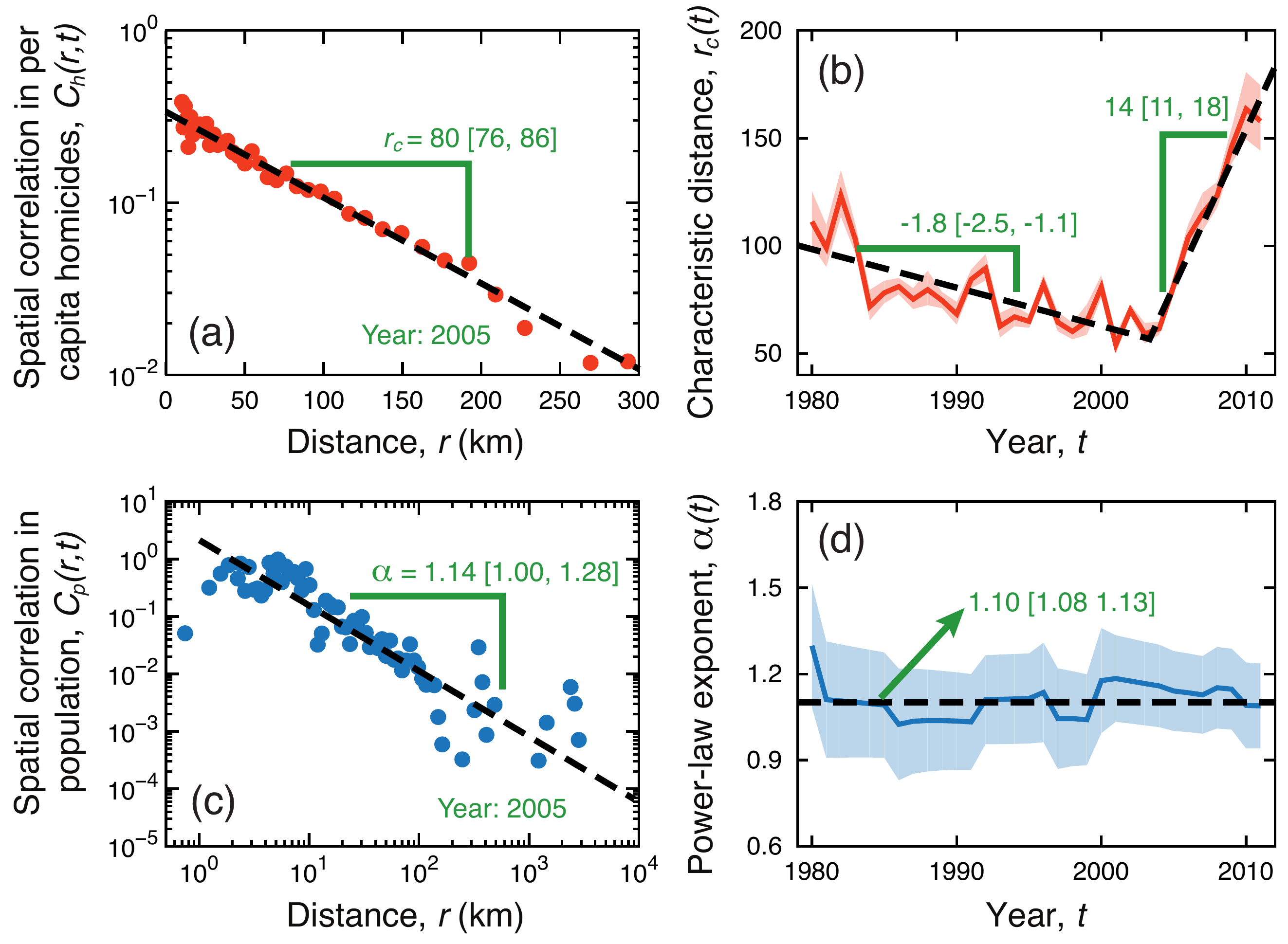}
\end{center}
\caption{Spatial correlations in the per capita homicide and in the population size. (a) Spatial dependence of correlation function $C_h(r,t)$ evaluated for the per capita homicides $h_i(t)$ in year $t=2005$. The red dots are the empirical values of $C_h(r,t)$ and the dashed line is an exponential fit (via ordinary least square method) to these data on mono-log scale. The value of the characteristic correlation length $r_c(t)$ and its 95$\%$ bootstrap confidence bounds are shown in the plot. (b) The red line shows the evolution of $r_c(t)$ along the years in our dataset. The shaded areas are 95$\%$ bootstrap confidence intervals for $r_c(t)$ and the two line segments are linear fits to values $r_c(t)$ versus $t$ in the periods of 1980 to 2003 and 2003 to 2011 (the linear coefficients and their confidence bounds are shown in plot). (c) Spatial dependence of the correlation function $C_p(r,t)$ evaluated for the population size $N_i(t)$ in year $t=2005$. The blue dots are the empirical values and the dashed line is a power-law fit to these data on log-log scale (the power-law exponent $\alpha$ is show in the plot). (d) The blue line illustrates the approximately constant behavior of $\alpha(t)$ over the years (shaded areas stand for 95$\%$ bootstrap confidence intervals), characterized by an average value of $1.10$.
}
\label{fig:2}
\end{figure*}

In order to start quantifying the spatial dynamics ruling $h_i(t)$, we evaluate the spacial correlation function 
\begin{equation}\label{eq:corr}
C_h(r,t) = \frac{\langle [h_i(t) - \mu(r,t)] [h_j(t) - \mu(r,t)] \rangle_{r_{ij}=r}}{{\sigma^2(r,t)}}\,,
\end{equation}
where $h_i(t)$ and $h_j(t)$ are the per capita homicides in the cities $i$ and $j$, $\mu(r,t)$ is the average and $\sigma^2(r,t)$ is the variance of the per capita homicides of cities separated by $r$ kilometers, and $\langle \dots \rangle_{r_{ij}=r}$ stands for the average over all pair of cities separated by $r$ kilometers. We have actually considered intervals of $r$ for evaluating $C_h(r,t)$ because of the natural discretization of our data. Figure~\ref{fig:2}(a) shows the spatial dependence of the correlation function $C_h(r,t)$ for the year $t=2005$ on mono-log scale. We observe that $C_h(r,t)$ is well approximated by an exponential decay spanning hundred of kilometers, that is, $C_h(r,t) \sim \exp{[ -r / r_c(t) ]}$, where $r_c(t)$ is the characteristic correlation length in year $t$. Similar exponential decays are observed for all years in our dataset. 
However, the value of $r_c(t)$ changes from year to year as depicted in Fig.~\ref{fig:2}(b). Between 1980 and 2003, $r_c(t)$ shows an approximately linear decreasing tendency of $1.8$~km by year, followed by an acute linear increasing trend of $14$~km by year in the period of 2003 to 2011. Our results thus indicate that value of the per capita homicide in a city has influence in its nearby cities through short-range correlations. However, it is worth noting that the values of $r_c(t)$ are comparable with the typical distances among Brazilian cities (average value of $\sim1000$~km); furthermore, the remarkable increasing trend of $r_c(t)$ over the past few years suggests that the influence/correlation that cities exert on each other has intensified. {It is very hard to point concrete origins for explaining this sharp variation in the values of $r_c(t)$ and, perhaps, the investigation of other urban indicators could help clarify this behavior. Homicides cases are actually correlated with several other urban indicators~\cite{Alves2,Alves4} including economic indicators such as income. These indicators also went through a period of quick variation over the last few years, which may have contributed to the changes in spatial dynamics of $h_i(t)$.}

{In addition of being correlated with other urban indicators, the evolutive features of $C_h(r,t)$ could also be  related to the spatial dynamics of the population size $N_i(t)$ due to the definition of $h_i(t)$.} To investigate this hypothesis, we evaluate the correlation function $C_p(r,t)$ considering the population size $N_i(t)$ in place of $h_i(t)$ in Eq.~\ref{eq:corr}. Figure~\ref{fig:2}(c) shows that the spatial correlation for the population size is well described by a power-law decay, $C_p(r,t)\sim r^{-\alpha(t)}$, with an exponent $\alpha(t)=1.14$ in the year $t=2005$. Analogous behaviors are obtained for all other years 
with practically the same power-law exponents. In fact, as shown in Fig.~\ref{fig:2}(d), the behavior of $\alpha(t)$ versus $t$ is well approximated by its overall average value ($\alpha(t) \approx 1.10$). Similar results were reported for the correlations in the population size of the United States cities~\cite{Makse3}. Thus, the evolution of $C_h(r,t)$ for the per capita homicides cannot be directly explained by the spatial dynamics of the population size.

\begin{figure*}[!ht]
\begin{center}
\includegraphics[scale=0.5]{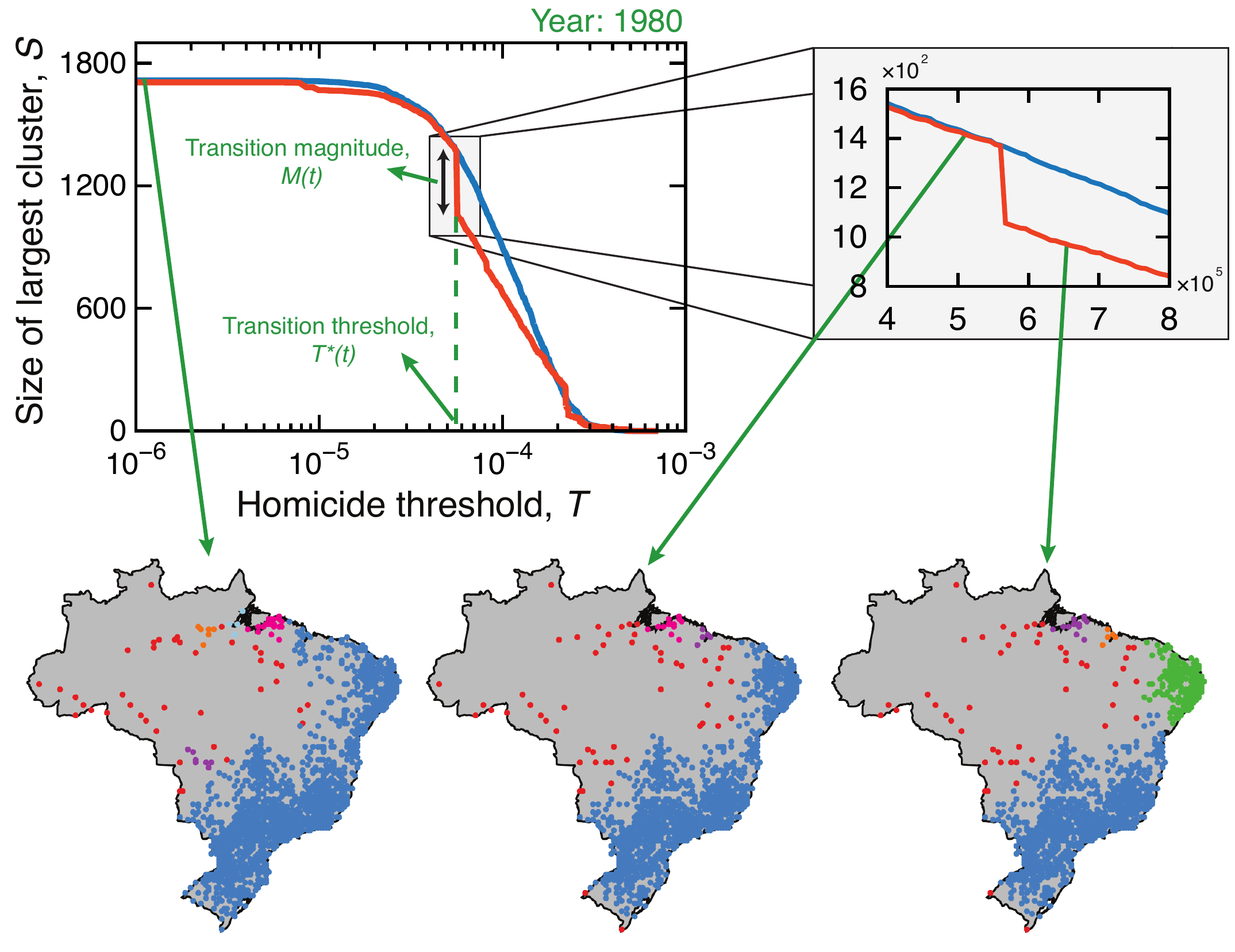}
\end{center}
\caption{Formation of spatial cluster of cities and a percolation-like analysis. The plot shows the size of the largest cluster $S$ (defined as the number of cities that belong to the cluster) as a function of the threshold $T$ in the per capita homicides. The red line shows the results considering data from the year $t=1980$ and the blue line shows the results after randomizing the values of the per capita homicides $h_i(t)$ among the cities. For the original data (red line), we observe that $S$ undergoes an abrupt change around a specific value of $T=T^{*}(t)$, followed by smaller subsequents abrupt changes. The inset shows a magnification of the region where this transition-like behavior occurs. The value of $T^{*}(t)$ is defined as the point where the jump in the $S$ is maximum and we have further defined the transition magnitude $M(t)$ as the size of this maximum jump (both $T^{*}(t)$ and $M(t)$ are illustrated in the plot). {The three maps illustrate the clusters identified by the DBSCAN algorithm for the values of $T$ indicated by the green arrows.}
}
\label{fig:3}
\end{figure*} 
{
Another interesting aspect of the exponential correlation in $h_i(t)$ is that it can be considered non-trivial in the context of the Edwards-Wilkinson equation~\cite{Edwards_Wilkinson}. This equation is usually employed for describing the stochastic kinetics of surfaces in the presence of diffusion and environmental randomness. In the continuum limit, we can write the Edwards-Wilkinson equation for the per capita homicides as 
\begin{equation}\label{eq_EW}
\frac{\partial }{\partial t}h({\bf r},t) = D\, \nabla^2 h({\bf r},t) + \eta({\bf r},t)\,,
\end{equation}
where $h({\bf r},t)$ represents per capita homicide in a point localized by the vector ${\bf r}$, $D$ is the diffusion coefficient, $\nabla^2$ is the Laplacian, and $\eta({\bf r},t)$ is an uncorrelated (in space and time) noise: $\langle \eta({\bf r},t) \eta({\bf r}',t') \rangle \propto \delta({\bf r} - {\bf r}')\delta(t - t')$. For this model, the correlation is usually obtained by taking Fourier transforms in space and time, evaluating the correlation in this double-transformed Fourier space and returning to the usual space via inverse Fourier transforms (see Refs.~\cite{Edwards_Wilkinson,Nattermann,Stanley} for more details). Following these procedures, we can write the spatial correlation in $d$ dimensions and for large values of $t$ as 
\begin{equation}
C_h(r)\sim 
\begin{cases}
r^{2-d} & \mbox{if }  d \neq 2 \\
 \ln(1/r) & \mbox{if } d = 2\,.
\end{cases}
\end{equation}
The previous result should be understood as a null model for our data and, thus, we would expect a very slow decay (a logarithm decay, since our system is essentially a two-dimensional one) for the spatial correlation in $h_i(t)$ assuming that the homicide dynamics is solely ruled by diffusive aspects and environmental randomness. In fact, an approximately logarithm decay for the spatial correlation was observed in some social systems (turnout rates in elections~\cite{Borghesi,Borghesi2} and obesity~\cite{Makse3}) and also for lightning activity rates~\cite{Ribeiro}. Naturally, the model of the eq.~\ref{eq_EW} is quite crude under the usual complexity involved in social systems; perhaps, the first ingredient to add to this model is a reaction term (leading us to a reaction-diffusion-like equation). Another important aspect of our system is the discrete and non-homogeneous spatial distribution of the cities, which may also play an important role on the spatial correlations. However, despite the importance of modeling, we believe that this modeling question may deserve a separated investigation.}

The spatial correlations in $h_i(t)$ also provide an indicative of the existence of spatial clusters of cities with similar values of per capita homicide. To address this question, we study the formation of spatial cluster of cities via a percolation-like analysis~\cite{Makse1,Makse2,Makse3}. Specifically, we define a homicide threshold $T$ and select all cities for which $h_i(t)\geq T$. For the set of cities satisfying this condition, we apply the density-based spatial clustering of applications with noise algorithm (DBSCAN --- as implemented in Python library Scikit-Learn~\cite{scikit}) for identifying possible spatial clusters. This algorithm finds statistically significant clusters and we have investigated the size of the largest cluster $S$ (that is, the number of cities in the largest cluster) as a function of the threshold $T$, as shown in Fig.~\ref{fig:3} (red line) for the year $t=1980$. Analogously to a phase transition, $S$ undergoes an abruptly change around a specific value of $T=T^{*}(t)$, where the largest cluster starts to be broken into smaller ones. { Similar behavior was observed for obesity rates in the United States~\cite{Makse3}. In our case,} the value of $T^{*}(t)$ is defined as the point where the jump in the $S$ is maximum and we have further defined the transition magnitude $M(t)$ as the size of this maximum jump. In this particular example, we have that at $T^{*}(t=1980)=5.67\times 10^{-5}$ the number of cities in the largest cluster decreases by $M(t=1980)=314$. We further observe the existence of smaller subsequents abrupt changes in $S$, producing a staircase-like behavior in $S$. In order to confirm that this transition-like behavior is not only related to the spatial location of the cities, we randomize the values of $h_i(t)$ among the cities and investigate again the relationship between $S$ and $T$. For the randomized data, Fig.~\ref{fig:3} (blue line) shows that the size of largest cluster continuously goes to zero as $T$ increases. Thus, the intricate behavior of $S$ versus $T$ observed for the original data cannot be directly related to the spatial distribution of the Brazilian cities. 

\begin{figure}[!ht]
\begin{center}
\includegraphics[scale=0.33]{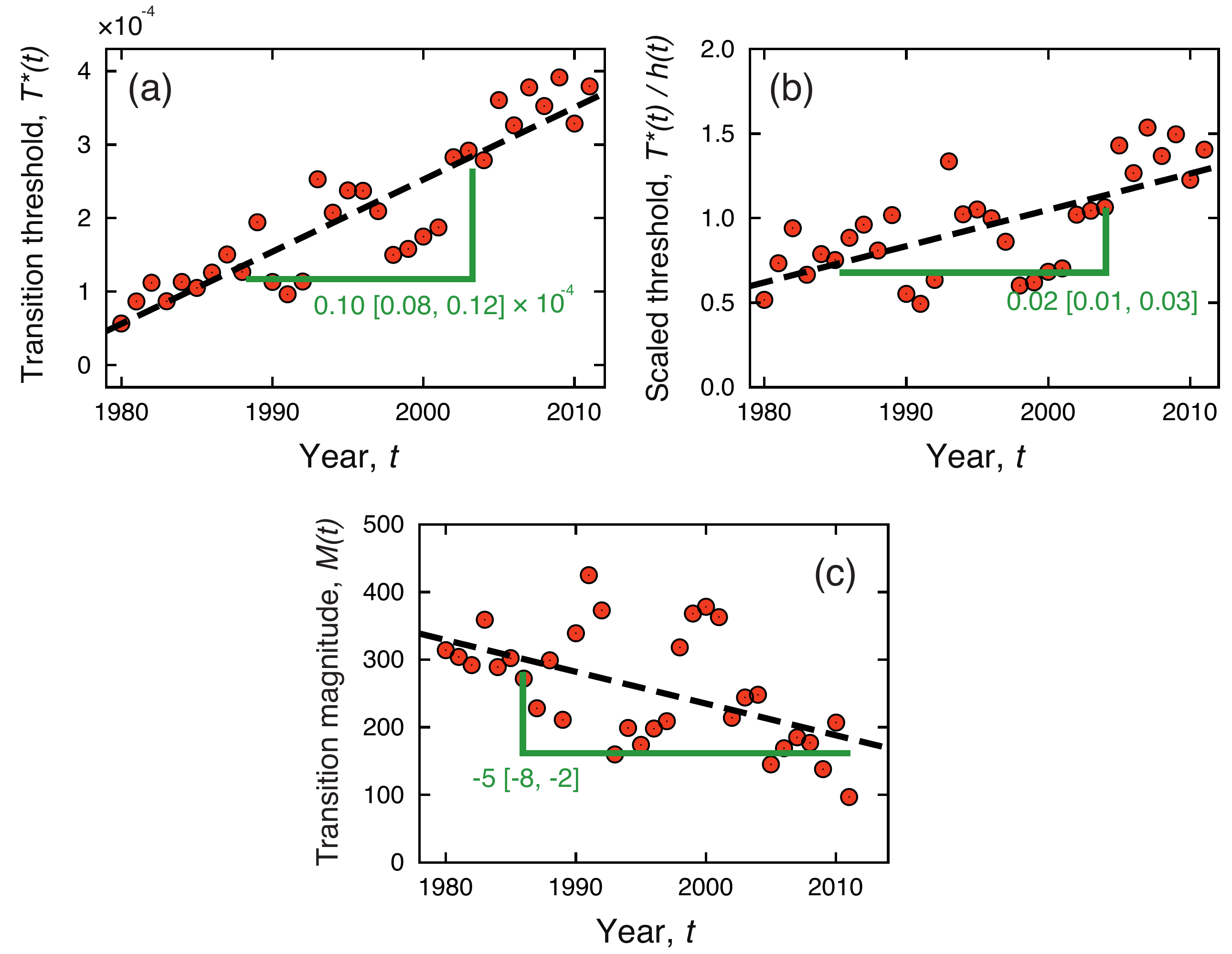}
\end{center}
\caption{
Evolutive aspects of the transition-like behavior. (a) Time evolution of the critical value $T^*(t)$ where the size of the largest cluster $S$ undergoes an abrupt change (red dots). The dashed line is a linear fit to these data and the linear coefficient as well as the confidence interval are shown in the plot. (b) Evolution of scaled homicide threshold $T^*(t)/h(t)$ (red dots). Again, we observe an increasing trend that is approximated by a linear function (dashed line) whose linear coefficient and the confidence interval are shown in plot. (c) Changes in the transition magnitude $M(t)$ over the years (red dots). Despite the noisy relationship, a statistically significant decreasing trend is observed, which can be approximated by a linear decay (dashed line) whose linear coefficient and the confidence interval are shown in plot.
}
\label{fig:4}
\end{figure}

We have applied the clustering analysis for all years in our dataset and a similar transition-like behavior is found for all of them, 
but with some evolutive features. The critical value $T^*(t)$ has increased over years with an approximately linear rate of $(0.10\pm0.01)\times10^{-4}$ per capita homicides by year, as shown in Fig.~\ref{fig:4}(a). Naturally, the growth of $T^*(t)$ should be related to the overall increasing trend in the per capita homicide of Brazil (Fig.~\ref{fig:1}). In order to account for this effect, we divide $T^*(t)$ by the overall per capita homicide in Brazil $h(t)$. Figure~\ref{fig:4}(b) shows that part of evolutive behavior of $T^*(t)$ is actually associated with $h(t)$; however, a statistically significant increasing trend is still observed for scaled threshold $T^*(t)/h(t)$. Even after scaling by $h(t)$, the scaled homicide threshold has almost tripled during the 31 years covered by our data. This result somehow agrees with the increasing in the characteristic correlation length $r_c(t)$, in the sense that the more intense correlations of recent years require larger thresholds $T^*(t)$ for breaking the spatial clusters. To further quantify the evolutive aspects of these transitions, we investigate whether the transition magnitude $M(t)$ has evolved along the years. Despite being a more noisy relationship, Fig.~\ref{fig:4}(c) shows that $M(t)$ has a significant decreasing trend that can be approximated by a linear decay of $\approx$5 cities per year. Again, we suspect that part of this behavior is related to the increasing in $r_c(t)$.

\section{Conclusions}
We have characterized the spatial dynamics of the homicide crimes in all Brazilian cities. Our results have shown that the per capita homicide in a city has influence in its nearby cities through short-range correlations. By investigating the evolution of characteristic correlation length, we verified that the influence/correlation among cities has considerably intensified over the latest years. Due the existence of these correlations, we have investigated the formation of spatial cluster of cities via a percolation-like analysis. Statistically significant clusters were observed and a phase transition-like behavior describing the size of the largest cluster as a function of the homicide threshold was also described. This transition-like behavior presents evolutive features characterized by an increasing in the homicide threshold where the transitions occur and by a decreasing in the transition magnitudes (length of the jumps in the cluster size). These two features seems to be related to the increasing of the characteristic correlation length, since more intense correlations require larger homicide thresholds for breaking the spatial clusters and may also contribute for the decreasing of the jumps in the cluster sizes. {Thus, our work sheds new lights on the spatial patterns of criminal activities at large scales and indicates that empirical patterns observed in the context of the ``broken windows theory'' at the city level (that is, the clustering of criminal activities) seems to emerge in a much larger scale. We further believe that our empirical findings open new possibilities for modeling criminal activities. Because of the successful description (at the city level), models based on reaction--diffusion equations and point processes together with extensions of the Edwards-Wilkinson equation are natural candidates to be tested for reproducing the patterns observed at at larger scales.}

\acknowledgements
We are all grateful to Capes, CNPq and Funda\c{c}\~ao Arauc\'aria for financial support. HVR thanks the financial support of the CNPq under grant number 440650/2014-3.

\bibliographystyle{plainnat}

\end{document}